\documentclass[9pt,twocolumn,twoside]{pnas-new}
\newcommand{\ball}{Squigglebot}
\usepackage{ulem}
\usepackage{multirow}

\setboolean{displaywatermark}{false}

\templatetype{pnasresearcharticle} 

\title{Role of kinematic constraints in the time reversal symmetry breaking of a model active matter}

\author[a]{Soumen Das}
\author[a]{Shankar Ghosh}
\author[b]{Tridib Sadhu}
\author[c,1]{Juliane U. Klamser}

\affil[a]{Department of Condensed Matter Physics and Material Sciences, Tata Institute of Fundamental Research, Mumbai 400005, India.}
\affil[b]{Department of Theoretical Physics, Tata Institute of Fundamental Research, Mumbai 400005, India.}
\affil[c]{Laboratoire Charles Coulomb (L2C), Université de Montpellier and CNRS (UMR 5221), 34095 Montpellier, France.}

\leadauthor{Das} 

\significancestatement{
Active matter breaks detailed balance at the microscopic level and generally cannot be described by the tools of equilibrium statistical mechanics. Developing a thermodynamic framework for active matter foremost requires understanding how dissipation drives active systems away from thermal equilibrium. However, it is often challenging to quantitatively characterize the signature of non-equilibrium physics. Conventional measures are typically invasive or require high-dimensional phase space information. The current work overcomes these limitations by proposing a non-invasive, experimentally viable measure of non-equilibriumness. Its effectiveness in quantifying dissipation is demonstrated in a novel tabletop artificial active system engineered for high-precision measurements of internal energetics, previously inaccessible in typical examples of active matter. In the process, the study uncovers a non-equilibrium fluctuation symmetry in dissipation, which is believed to be a universal characteristic of active energetics. These quantifications of dissipation open new avenues for probing self-organization principles in active matter.
} 

\authorcontributions{SD and SG designed and performed the experiments. SD, SG, and JK analyzed the data. JK conducted computer simulations and generated the final figures. SG, JK, and TS wrote the manuscript.}
\authordeclaration{There are no competing interests.}
\correspondingauthor{\textsuperscript{1} Juliane U. Klamser: juliane.klamser@umontpellier.fr }


\begin{abstract}
Active-matter systems are inherently out-of-equilibrium and perform mechanical work by utilizing their internal energy sources. Breakdown of time-reversal symmetry (BTRS) is a hallmark of such dissipative non-equilibrium dynamics. We introduce a robust, experimentally accessible, noninvasive, quantitative measure of BTRS in terms of the Kullback-Leibler divergence in collision events, demonstrated in our novel artificial active matter, comprised of battery-powered spherical rolling robots whose energetics in different modes of motion can be measured with high precision. Our dimensionless measure characterizes how dissipation and internal energetics are influenced by kinematic constraints from interactions with the environment. We propose this measure of BTRS as an empirical estimate of the distance from equilibrium. An energetic insight into this departure of active matter from equilibrium comes from our demonstration of a non-trivial fluctuation symmetry, which reveals a potentially universal thermodynamic characteristic of active energetics. As a many-body consequence of BTRS in our experimental active system, we demonstrate the emergence of activity-induced herding, which has no equilibrium analogue.
\end{abstract}

\dates{This manuscript was compiled on \today}

\begin{document}
\maketitle
\thispagestyle{firststyle}
\ifthenelse{\boolean{shortarticle}}{\ifthenelse{\boolean{singlecolumn}}{\abscontentformatted}{\abscontent}}{}

\dropcap{U}nlike inert objects, living matter converts energy from its internal sources for locomotive motion. Such systems, generally termed as active matter \cite{Bowick2022,ramaswamy2017active,ramaswamy2010mechanics_annurev-conmatphys}, are self-driven out of equilibrium at the individual constituent level. The driving stems from the conversion of some form of stored energy (either onboard or in the ambient environment), which is then dissipated to move systematically against friction with the surrounding medium. For instance, living systems can harness energy at the molecular scale through ATP hydrolysis and dissipate it across larger spatio-temporal scales, leading to the emergence of self-organized structures \cite{agrawal2017chromatin,ndlec1997self,juelicher2007active,ballerini2008interaction}. Synthetic active matter, such as Janus particles \cite{vicsek1995novel} and driven granular systems \cite{deseigne2010collective}, are independently driven by the surrounding medium, injecting energy at the microscopic scale through individual constituents. What unites these examples is the ability of active components to self-regulate their energy consumption, distinguishing them fundamentally from externally driven non-equilibrium systems like sheared fluids. This sustained consumption and dissipation of energy at the microscopic level transcends the constraints imposed by the Stokes-Einstein Fluctuation Dissipation Theorem (FDT), thus breaking the statistical time reversibility of equilibrium dynamics.

In recent times, there has been a surge of interest in quantifying the breakdown of time-reversal symmetry (BTRS) as a natural measure of the distance from equilibrium and in understanding its relation to energy dissipation. The violation of FDT has been a traditional method for detecting BTRS  \cite{PhysRevE.77.051111,C0SM01484B,Levis2015}. However, the information extracted from this approach is rather limited and requires direct perturbations for measuring response properties, which are invasive and challenging in practice \cite{Mizuno2007}. In comparison, stochastic thermodynamics offers more elaborate tools to quantify non-equilibriumness \cite{Seifert2005,Mallick2018}. One such non-invasive measure is in terms of the entropy production, which probes the BTRS at the microscopic trajectory level. Attempts have been made \cite{Fodor2016PRL,pruessner2022,OByrne2022a} to utilize the entropy production rate to quantify how these active systems depart from equilibrium. However, estimating the entropy production in practice remains a formidable task due to the challenges in sampling the entire phase space trajectories. These computations are challenging, particularly outside of simplified theoretical models, and even more so experimentally. The challenge is intensified for multi-scale active matter, where kinematic constraints resulting from interactions between many particles and with the environment are known to influence both the dynamics and the steady state \cite{Murali2022PRR,2015_Solon}. Our current work seeks to address these limitations by proposing alternative measures of BTRS and testing them in a novel experimental realization of artificial active matter.

The objectives of this work are two-fold: introducing a simple, non-invasive, local, dynamical measure of BTRS and demonstrating the effect of environmental constraints on it through direct measurements in an experimental active matter system. Our quantitative measure of BTRS is defined in terms of the Kullback-Leibler divergence \cite{Kullback59,Cover2006} (also known as the relative entropy) in the distribution of scattering angles in binary collisions. We demonstrate that local dynamical measurements of this quantity can reveal the broken time-reversibility without access to net drifts or flows in phase space. Importantly, unlike the Markovian assumption in the definition of entropy production \cite{Seifert2005,Mallick2018}, our new measure does not rely on building theoretical models.

An emphasis of our study is on the influence of environmental constraints on the BTRS and dissipation. The interaction of active matter with dynamically constrained structures has been exploited in the construction of active engines, with a focus on efficiency \cite{Lee2022PRE}. However, little is known about the thermodynamic aspects of the process by which an individual active particle converts its internal energy to achieve desired locomotion. A typical extended active particle possesses multiple internal modes coupled to its internal energy source, which it self-regulates in response to interactions with the environment. As a result, environmental constraints can impact the energy expenditures of these internal modes, thereby affecting the BTRS. The influence of the environment on the energetics of individual active constituents has been largely overlooked \cite{Bowick2022}, mainly due to challenges in accessing the internal modes of experimental active systems. We overcome this limitation by engineering an artificial active particle, which we call the "Squigglebot". The Squigglebot's energy intake, as well as its energy expenditure into different modes of locomotion, can be monitored with high precision. This allows us to experimentally compare our dynamical measure of BTRS with energy expenditure, and how they change under different environmental constraints. We demonstrate that our dynamical measure from collision events is much more efficient in quantifying BTRS than simple measures of dissipation in the energy consumption of internal modes within the active particle. 

An alternative energetic measure of BTRS comes from a deeper underlying structure of the internal modes of an active particle. The unprecedented access in our experiment to the energy expenditure of internal modes unravels a type of Gallavotti-Cohen-Evans-Morriss (GCEM) fluctuation symmetry \cite{Evans1993PRL,Gallavotti1995PRL,Kurchan1998JPA,Lebowitz1999JSP} in active energetics. This discovery is surprising given the athermal nature of the system. This non-trivial fluctuation symmetry identifies a potentially universal thermodynamic characteristic of active energetics. Moreover, a dimensionless parameter in the fluctuation symmetry, analogous to the entropy production rate \cite{Fodor2016PRL}, offers a quantitative estimate of BTRS. Our measurement corroborates a similar influence of kinematic constraints on active dynamics as inferred from the collision events: BTRS is not inherent to individual active particles; rather it is strongly influenced by environmental constraints, and suitably chosen constraints can effectively bring an active particle closer to equilibrium. 

The BTRS liberates active dynamics from the thermodynamic constraints of equilibrium, often resulting in novel many-body self-organization  that lacks direct equilibrium analogues. Such cooperative phenomena are one of the reasons behind the intense interest in active matter. Motility-induced phase separation (MIPS) is one well-known \cite{Tailleur_Cates} collective phenomenon where dense clusters form solely due to the self-propulsion of particles, without the need for attractive pair interactions. In our experimental active system, we demonstrate a similar many-body consequence of BTRS, wherein a single active particle causes a herding effect in the surrounding passive medium, inducing macroscopic density inhomogeneity. This phenomenon is one of the simplest manifestations of activity-induced macroscopic effects that is realizable in finite systems.

We organize the article in the following order. We begin by introducing our system of controllable artificial active matter and quantifying its single-particle characteristics in Section~\ref{S:ExpSetup}. In the following Section~\ref{sec:btrs}, we introduce our new dynamical measure of BTRS and report its observed values in our experiment subjected to different kinematic constraints. In Section~\ref{sec:fdt}, we demonstrate a thermodynamic characteristic of active energetics in terms of the GCEM fluctuation symmetry and use it as an alternative energetic measure of BTRS. In Section~\ref{sec:passive medium}, we discuss activity-induced herding and conclude in Section~\ref{sec:Conclusion} with open directions.

\begin{figure*}
		\centering
		\includegraphics{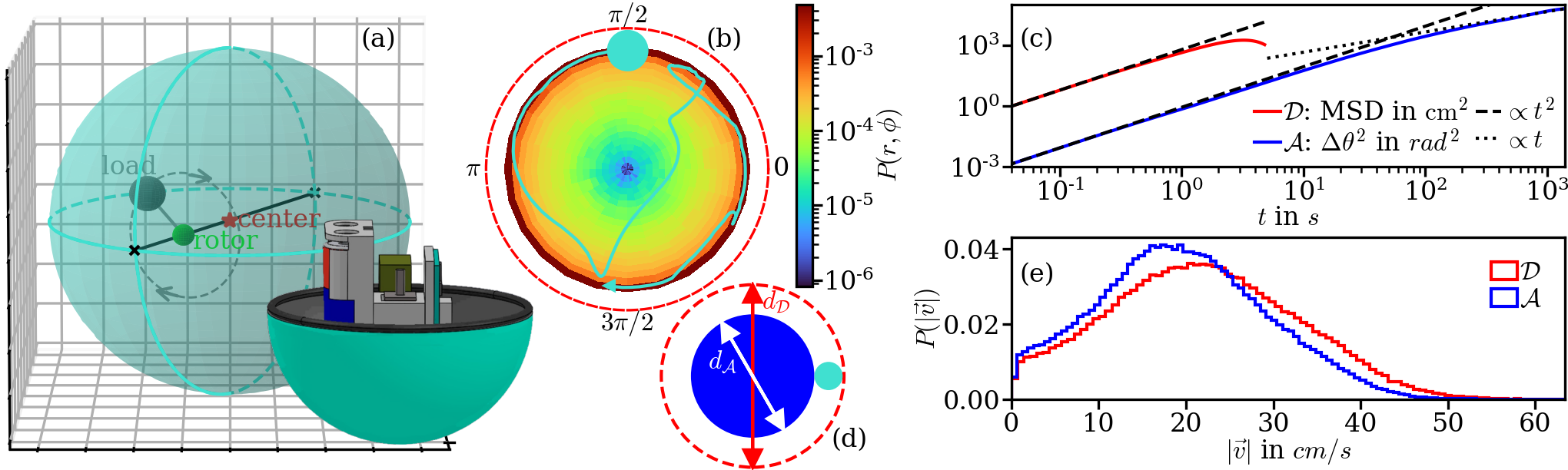}
		\caption{(a) \textbf{Schematic of the active particle:} Spherical shell (turquoise  sphere) and stator (black line) form a single rigid body. The rotor (green sphere) driven by a battery-powered motor (not shown) and the load (gray sphere) form another rigid body. When the load loops around the stator a torque is exerted on the latter and thus directly transferred to the shell of the particle resulting it to roll in a persistent manner. The schematic in the front shows the battery and motor assembly (details are in the Materials and Methods). (b) Probability distribution of the position (see color bar) of the \ball{} (turquoise circle) in a disc shaped arena $\mathcal{D}$ (the boundary indicated by red dashed line) and a typical particle trajectory (turquoise line). (c) Mean square displacement (red line) of the \ball{} in $\mathcal{D}$ showing a ballistic behaviour (MSD$\propto t^2$) and a cutoff at $t\approx 3$ resulting from the finite size of the arena. Mean square angular displacement (blue line) for the \ball{} confined in an annulus shaped arena $\mathcal{A}$ shown in (d). The diameter of the outer boundary $d_\mathcal{D}=480\rm{mm}$ and the diameter of the inner boundary $d_\mathcal{A}=320\rm{mm}$. (e) Probability distribution of the speed in the two arenas $\mathcal{D}$ and $\mathcal{A}$. \label{fig:ballIntro}}
\end{figure*}

\section{The experimental system: \ball{} \label{S:ExpSetup}}
We present the \ball{} as an experimental active particle based on a spherical cat toy called the Squiggle Ball\textsuperscript{\texttrademark}.  Fig.~\ref{fig:ballIntro}a shows a schematic representation of the \ball{} which consists of a spherical plastic shell (turquoise sphere) of diameter $80\;\rm{mm}$, to which one stator (black line) is rigidly fixed diametrically. Therefore, the stator and shell represent one rigid body that cannot move independently from each other. The motion of the \ball{} is generated by a battery-powered motor that drives a rotor, to which a load is attached rigidly at some distance from the stator (see Fig.~\ref{fig:ballIntro}a). When the load loops around the stator, a torque is exerted on the stator and thus directly transferred to the shell of the \ball{}, causing the ball to roll in the direction perpendicular to the stator. However, as the rotor is positioned off-center on the stator, the inertia of the load causes the stator to  wag, resulting the \ball{} not rolling smoothly but squiggling (see \cite{Gfroerer2007,Prentis2000,Riedl2023} for related works). The instabilities due to the squiggle can initiate changes in the direction of motion, either via smooth curves or tumbling events that cause drastic changes in direction. Tumbling can lead to an almost instantaneous inversion of the direction of motion. Overall, the \ball{} exhibits systematic persistent motion characteristic of active particles, with a well-defined persistence time and length.

Fig.~\ref{fig:ballIntro}b shows a typical particle trajectory in a disc shaped arena $\mathcal{D}$, along with the probability distribution of particle's position. The \ball{} spends most of its time moving along the boundary wall and occasionally detaches from it to explore the bulk of the arena. As shown in Fig.~\ref{fig:ballIntro}c, the resulting mean square displacement grows ballistically ($\rm{MSD}\propto t^2$), but its maximum value is limited by the size of the arena. The persistence length of the \ball{} is larger then diameter of the disc arena $\mathcal{D}$, preventing us from capturing the crossover from ballistic to diffusive motion, which is characteristic of active particles. This limitation can be overcome by confining the \ball{} in an annulus-shaped arena $\mathcal{A}$, shown in Fig.~\ref{fig:ballIntro}d, which constrains the particle to move in a circle. In polar coordinates, the position of the \ball{} $\vec{r} = (r,\theta)$, where $r$ can be considered constant. We record the angle by adding increments $\Delta \theta(t)$ in small $\Delta t$ time intervals such that $\theta(t+\Delta t) = \theta(t) + \Delta \theta(t)$, and the angle takes values in the range $-\infty < \theta < \infty$. This allows us to compute the mean square angular displacement of the \ball{} as $\langle \Delta \theta^2 (t)\rangle = \langle (\theta(t+t_0)-\theta(t_0))^2\rangle$ for arbitrary $t_0$, which is also shown in Fig.~\ref{fig:ballIntro}c. The mean square angular displacement initially shows ballistic scaling, but we are now able to capture the crossover to diffusive motion at $t\approx 50\rm{s}$, which is an upper limit for the free-particle persistence time of the \ball{}. This is because the \ball{} tends to align its direction of motion along the confining wall (see Fig.~\ref{fig:ballIntro}b), effectively increasing the persistence time compared to a particle freely moving without kinematic constraints.

We stress that no direct comparison of the particle speed $v = |\vec{v}|$ in the two arenas can be made from the mean square displacements in Fig.~\ref{fig:ballIntro}c. Instead, we provide in Fig.~\ref{fig:ballIntro}e the full distribution of $v$ to show that (i) the particle speed is not a constant but widely distributed with a well-defined mean ($\bar{v}_\mathcal{D} \approx 21.94$, $\bar{v}_\mathcal{A} \approx 19.55$) and (ii) the \ball{} confined in the annulus-shaped arena moves slightly slower, despite larger persistence time, than in the less constrained disk-shaped arena, with a relative difference $(\bar{v}_\mathcal{D}-\bar{v}_\mathcal{A})/(\bar{v}_\mathcal{D}+\bar{v}_\mathcal{A})$ of less than $6\%$.

\section{A general, non-invasive measure for BTRS}\label{sec:btrs}
A cornerstone of stochastic thermodynamics is the quantification of BTRS as the Kullback-Liebler divergence between the probability of forward and time-reversed trajectories. This information-theoretic measure of distinguishablity between distributions is also known as the relative entropy \cite{Maes_2003,Parrondo2009}. As dissipation is correlated with statistical irreversibility, the relative entropy has been proposed as an equivalent quantitative estimator for dissipation. Specifically, the relative entropy provides \cite{Seifert2005} a lower bound on the amount of energy dissipated to drive the system out of equilibrium.

The irreversibility measure of the relative entropy has proven to be a robust experimental and computational tool to detect non-equilibrium even in the absence of observable flow. However, such a quantity is challenging to measure accurately as it requires large amounts of data of the entire phase space trajectories, especially when there is no observable current \cite{Roldan2010}. Simpler measures can be constructed with the relative entropy involving suitable dynamical events with reduced dimensionality. We propose one such experimentally accessible measure and demonstrate its efficacy in quantifying BTRS. (For an alternative simple measure see \cite{Mori2021}.)
    
\begin{figure}[H]
		\centering
  \includegraphics[width=0.75\linewidth]{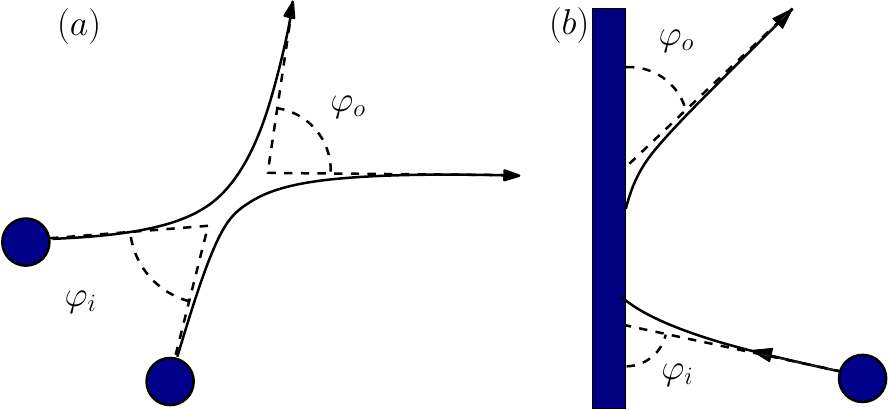}
		\caption{(a) Schematic of a binary collision event between two particles with incoming angle $\varphi_i$ and outgoing angle $\varphi_o$. (b) Similar dynamical interaction between a particle with a stationary wall.}
		\label{fig:collision}
\end{figure}

Consider binary collision events between particles with incoming angle $\varphi_i$ and outgoing angle $\varphi_o$, as sketched in Fig.~\ref{fig:collision}a. A similar construction applies for the incidence angle $\varphi_i$ of a particle onto a boundary wall and the outgoing angle $\varphi_o$ with which the particle leaves the wall. If the particles are confined in a finite domain, the distributions $P(\varphi_i)$ and $P(\varphi_o)$ will reach stationarity over time. The relative entropy or the Kullaback-Leibler divergence \cite{Kullback59,Cover2006} of these stationary distributions is
\begin{equation}\label{eq:KLD}
    {\rm KLd}=\int d\varphi P(\varphi_o=\varphi)\log\frac{P(\varphi_o=\varphi)}{P(\varphi_i=\varphi)}.
\end{equation}
This quantity is positive, attaining minimum (${\rm KLd} = 0$) when the two distributions are identical. This occurs when time reversal symmetry is preserved, i.e., in an equilibrium state \cite{Mallick2018}. When time reversal symmetry is broken, the two distributions differ, and the KLd monotonically grows with the difference, thus providing a quantitative estimator of BTRS. 

The KLd in \eqref{eq:KLD} can be thought of as a coarse-grained version of the KLd of distribution of phase space trajectories and their time-reversed trajectories, which is a standard measure of the entropy production \cite{Seifert2012,Mallick2018}. Compared to the definition in \eqref{eq:KLD}, where the measurement is done only at the collision events, the KLd of trajectories involves measurements at every instance of evolution. The latter contains higher dimensional information and, according to stochastic thermodynamics \cite{Seifert2005}, provides a lower bound on the energy dissipation. It has been experimentally established \cite{tan2021scaledependent} that a KLd constructed from less frequent measurements still provides a lower bound on dissipation, although weaker. Similarly, we propose that the irreversibility measure in \eqref{eq:KLD} offers an estimation of dissipation; the crudeness of the estimate is compensated by the simplicity of its measurement.

One fundamental difference between the measure in \eqref{eq:KLD} and  entropy production is that the latter is a time-integrated quantity measured over long trajectories, while the former involves stationary statistics.

We demonstrate this new measure \eqref{eq:KLD} of BTRS in the \ball{}-system under different kinematic constraints. First, we consider the case of a single \ball{} confined in the disk-shaped arena (see Fig.~\ref{fig:ballIntro}b). As the position of the confining wall does not change over time, we call this constraint scleronomic. Second, we consider binary collisions between two particles moving in the disk-shaped arena and compare the degree of non-equilibrium when binary collisions occur in the bulk versus those that occur when particles are in contact with the wall. For binary collisions, the motion of one particle is constrained by the position of the second particle, which explicitly depends on time; thus, we call this constraint rheonomic.

\subsection{Scleronomic constraint}	\label{sec:scleronomic}
\begin{figure}[t]
		\centering
  \includegraphics[width=1.\linewidth]{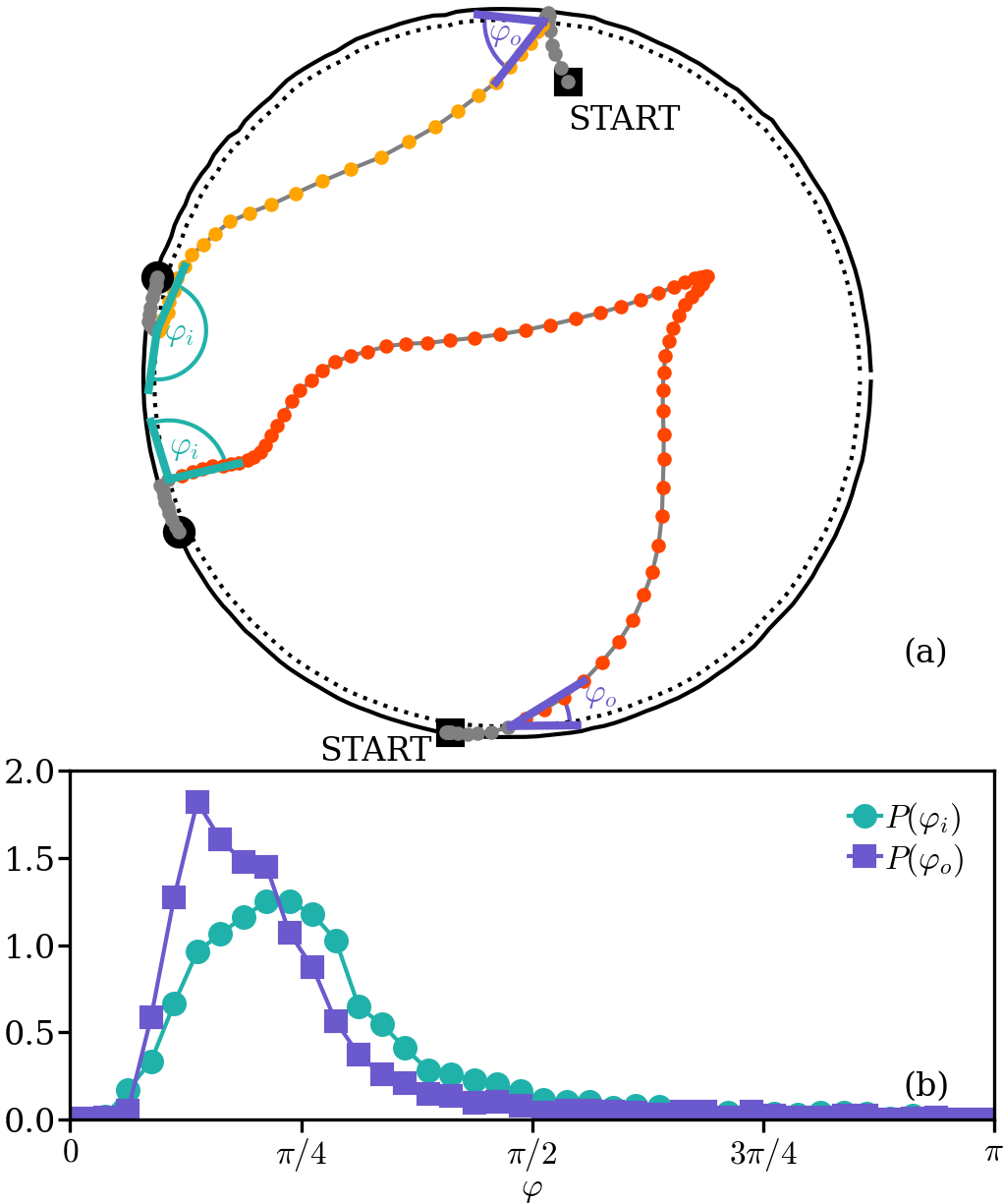}
        \caption{(a) Two bulk trajectories of a single \ball{} leaving the boundary wall at an angle $\varphi_o$ and colliding with the wall at an incidence angle $\varphi_i$. The circular solid black line indicates the center position of the \ball{} when in contact with the wall. We define that a bulk trajectory leaves (or joins) the wall when the \ball{} crosses the circular dotted line, which is positioned at one-tenth of the particle's diameter away from the wall-contact line. (b) Probability distribution of $\varphi_o$ (squares) and $\varphi_i$ (circles) sampled over a large number ($\approx 3500$) of bulk trajectories each lasting longer than $2$~seconds.}
		\label{fig:btrs measure single ball}
  		\centering
\end{figure}

For this scenario of a single \ball{} inside the disc arena $\mathcal{D}$, full bulk trajectories are shown in Fig.~\ref{fig:btrs measure single ball}a, where a trajectory starts when the ball leaves the wall with an outgoing angle $\varphi_o$ and ends when the ball joins the wall with an incoming angle $\varphi_i$. The angles $\varphi_i$ and $\varphi_o$, in the range $[0,\pi]$, are measured with respect to the tangent of the circular boundary wall, following a convention (clockwise or counterclockwise) such that the angular arc avoids the trajectory immediately after and before the collision, respectively. Schematics of incoming and outgoing angles are shown in Fig.~\ref{fig:btrs measure single ball}a. An acute incidence (outgoing) angle corresponds to incoming (outgoing) trajectories that, upon collision with the boundary, keep moving with a gradual change in direction, while an obtuse angle is associated with a sudden reversal or tumbling of self-propulsion direction. This convention of angles ensures that for a time-reversed trajectory, incoming and outgoing angles are interchanged, and their statistics are identical in equilibrium.

Fig.~\ref{fig:btrs measure single ball}b shows the probability distributions $P(\varphi_o)$ and $P(\varphi_i)$, both of which are peaked around $\varphi_i=\pi/4$, with their probability smoothly decaying around $0$ and $\pi$. This indicates that the \ball{} typically approaches the wall from the bulk at an acute grazing  angle and, without tumbling, either gets reflected away from the wall or aligns its self-propulsion direction parallel to the wall, resulting in a crawling motion along the wall. This tendency of the ball to crawl against the wall was observed in Fig.~\ref{fig:ballIntro}. For a crawling trajectory to lift off the boundary wall, the \ball{} must actively rotate its self-propulsion direction away from the wall. This can either happen smoothly (associated with an acute $\varphi_o$) or through a tumbling event (associated with an obtuse $\varphi_o$) where the \ball{} reorients its propulsion direction by flipping the stator axis with the load. The low probability of obtuse angles in Fig.~\ref{fig:btrs measure single ball}b shows that such tumbling events are rather infrequent.

Even though the two distributions look alike, the $P(\varphi_o)$ is peaked at a smaller angle than the peak-position of $P(\varphi_i)$. The former distribution is also narrower compared to the latter. This characteristic difference between the two distributions signals BTRS, which is quantified by the non-zero value $KLd \approx 1.545$ of the irreversibility measure in \eqref{eq:KLD}. For a reversible scenario, such as for passive particles, there is no difference in the statistics between a forward and its time-reversed trajectories, and therefore $KLd=0$.

\subsection{Rheonomic constraint}
\begin{figure}
\centering
\includegraphics[width = 1.\linewidth]{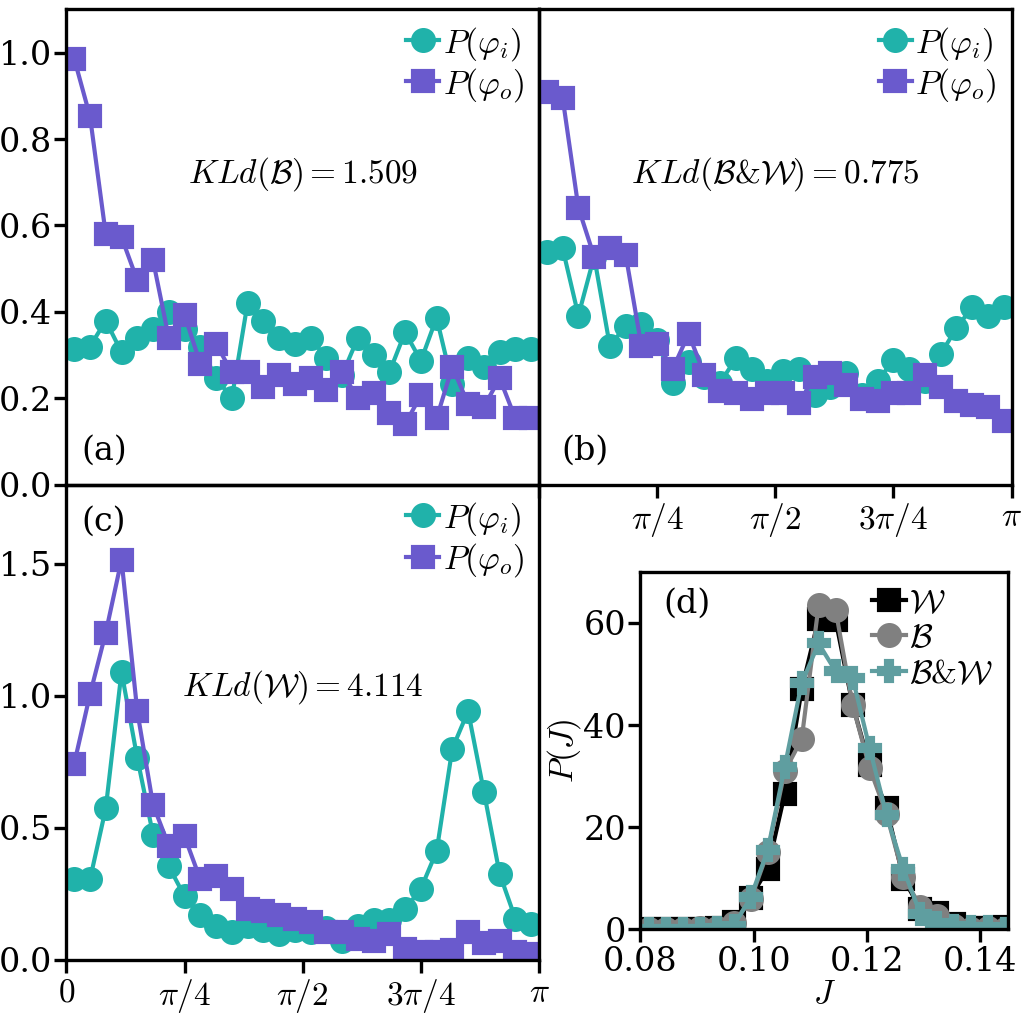}
\caption{(a-c) Probability distribution of the $\varphi_i$ and $\varphi_o$ for two-particle collisions for the cases (a) both particles are in the bulk ($\mathcal{B}$) at the time of collision $t_c$, (b) one particle is in contact with the boundary wall ($\mathcal{B}\text{\&}\mathcal{W}$), and (c) both particles are in contact with the wall ($\mathcal{W}$). Corresponding KLd of collisions is also indicated. (d) For these three cases, the probability distribution of the time averaged current $J$ in a short time interval of $2s$ around collisions.
}
\label{fig:two ball}
\end{figure}

The second scenario of kinematic constraint that we consider involves the dynamics of two \ball{}s moving in the disc arena. The two balls interact with each other through hard-core repulsion when they occasionally collide.   

There are three collision scenarios based on kinematic constraints at the moment of collision: (i) both \ball{}s are in the bulk, (ii) one of the \ball{} is in contact with the boundary wall, and (iii) both \ball{}s are in contact with the wall. These collision events are henceforth labeled as bulk, bulk-boundary, and boundary collisions, respectively. 
 
The distributions of the collision angles shown in Figs.~\ref{fig:two ball} convey interesting differences between the three scenarios. When the collision occurs in the bulk, the distribution of the incidence angle $\varphi_i$, shown in Fig.~\ref{fig:two ball}a, is relatively flat, compared to the distribution of the outgoing angle $\varphi_o$, which is peaked at small angles. This reflects the tendency of two persistently moving \ball{}s to travel parallel to each other after the collision, despite approaching uniformly from all angles. This preferential alignment of self-propulsion direction is analogous to the crawling of \ball{}s against the boundary wall noted in Fig.~\ref{fig:btrs measure single ball}. 

For bulk-boundary collisions in Fig.~\ref{fig:two ball}b, the distribution of $\varphi_o$ is similar to the corresponding distribution in pure bulk-collisions. In contrast, the distribution of $\varphi_i$ is morphed into an U-shape. The higher probability of low incidence angles is due to events where a \ball{} coming from the bulk collides with a \ball{} at the boundary from behind, while large incidence angles are due to near head-on collisions. In both cases, the \ball{} approaches the boundary from the bulk at low grazing angles as seen in Fig.~\ref{fig:btrs measure single ball}. 

The difference in collision statistics is more visible for pure boundary-collisions, as shown in Fig.~\ref{fig:two ball}(c). When both colliding balls are touching the boundary, they could approach at small (head-on collision) or large (chase-and-run collision) incidence angles. This is evident in the double-peak structure of distribution of $\varphi_i$. Strikingly, the distribution of the outgoing angle is uni-modal. The absence of a second peak at large outgoing angles shows that after most collisions, including the head-on scenario, the colliding balls tend to move in the same direction after the collision. This is due to the lack of momentum conservation in the \ball{} dynamics.

The difference in the distribution of the collisions angles in the three scenarios, captures the effect of kinematic constraints. The numerical value of the KLd in \eqref{eq:KLD} for collision angles, shown in the Figs.~\ref{fig:two ball} and summarized in Table.~\ref{tab:KLd}, gives a quantitative estimate of this effect of constraint on BTRS. The KLd is smallest for bulk-boundary collisions and largest for the pure boundary collisions, in accordance with the visual difference of the distributions in their collision angles.
Even though, in the boundary and bulk-boundary collisions, there are additional constraints compared to bulk-collisions, KLd is larger in one case and lower for another, emphasizing the importance of the details of constraints. 

The effectiveness of KLd in quantifying BTRS is more evident when compared with energy expenditure. A direct measure of energy consumption is the current drawn from the motor for propelling the \ball{} by converting its internal rotational motion into translational motion. A rise in motor power consumption indicates higher dissipation, hence larger BTRS. The distribution of currents, drawn over a small time interval around the collisions, is shown in Fig.~\ref{fig:two ball}(d). It is difficult to discern any significant difference between the distributions for different collision events, even in their peak positions, which are indicative of the average current. This demonstrates that, while the KLd of collision events efficiently quantifies BTRS, the mere distribution of energy current around collision events proves ineffective. In the following section, we show that an effective energetic measure of BTRS comes from a non-trivial symmetry in the current statistics. 

\begin{table}
\centering
\caption{A comparison of the KLd in \eqref{eq:KLD} under different kinematic constraints. \label{tab:KLd}}
\begin{tabular}{|c|l|ccc|}
\hline
\multirow{2}{*}{Constraint} & \multirow{2}{*}{Scleronomic} & \multicolumn{3}{c|}{Rheonomic}                                                \\ \cline{3-5} 
                            &                              & \multicolumn{1}{c|}{bulk}  & \multicolumn{1}{c|}{bulk \& boundary} & boundary \\ \hline
KLd                         & \multicolumn{1}{c|}{$1.545$} & \multicolumn{1}{l|}{$1.509$} & \multicolumn{1}{c|}{$0.775$}            & $4.114$    \\ \hline
\end{tabular}
\end{table}

\section{Fluctuation symmetry as a measure for BTRS}\label{sec:fdt}
The fundamental reason behind the time-reversal symmetry breaking (BTRS) in active matter is the breakdown of the fluctuation-dissipation relation for the internal energetics of individual constituents. The BTRS measured in the Kullback-Leibler divergence (KLd) in \eqref{eq:KLD} of collision events is a consequence of this internal irreversibility.

Even though time-reversibility is broken outside equilibrium, under most scenarios, it is broken in a specific manner \cite{Mallick2018}. This remarkable regularity of nature was discovered as a symmetry in the distribution of entropy production (identified with phase space contraction) and popularly known as the Gallavotti-Cohen-Evans-Morris (GCEM) fluctuation relation. It was first put forward by Evans \textit{et.al.} \cite{Evans1993PRL} and subsequently proven for chaotic dynamical systems \cite{Gallavotti1995PRL} and Markov processes \cite{Kurchan1998JPA,Lebowitz1999JSP}. This nontrivial fluctuation symmetry provides a relative weight for the transient violation of the second law of thermodynamics \cite{Seifert2005,Mallick2018}, and has been experimentally verified  \cite{Ciliberto2010JSM,Menon_2004}.

A version of the GCEM relation is a fluctuation symmetry in the statistics of energy transport between two thermal reservoirs at different temperatures  \cite{Derrida2004JSP,Derrida2007JSTAT,Dhar2008AP,Barato2012}.
One might think of internal energetics of an active particle in analogy with the above energy transport from an inbuilt source of energy to locomotive degrees of freedom. In this analogy, if $Q_T$ is the net energy flux drawn from the internal energy source over a duration $T$, then the probability $P(Q_T/T=j_T)$ would follow the symmetry
\begin{equation}
\frac{P(j_T)}{P(-j_T)}\sim e^{s T j_T},\label{eq:gc s}
\end{equation}
for large $T$.
In an equilibrium state, $P(j_T)$ is an even function and therefore $s=0$. A non-zero value of $s$ reflects the breakdown of time-reversal symmetry due to non-vanishing energy currents. Therefore, $s$ offers a quantitative measure of BTRS.
In fact, for energy transport between thermal reservoirs at temperatures $T_h$ and $T_c$, $s=\frac{1}{T_c}-\frac{1}{T_h}$ and $s j_T$ is the entropy production rate \cite{Seifert2012,Derrida2007JSTAT}.
\begin{figure}[t]
\centering
\includegraphics{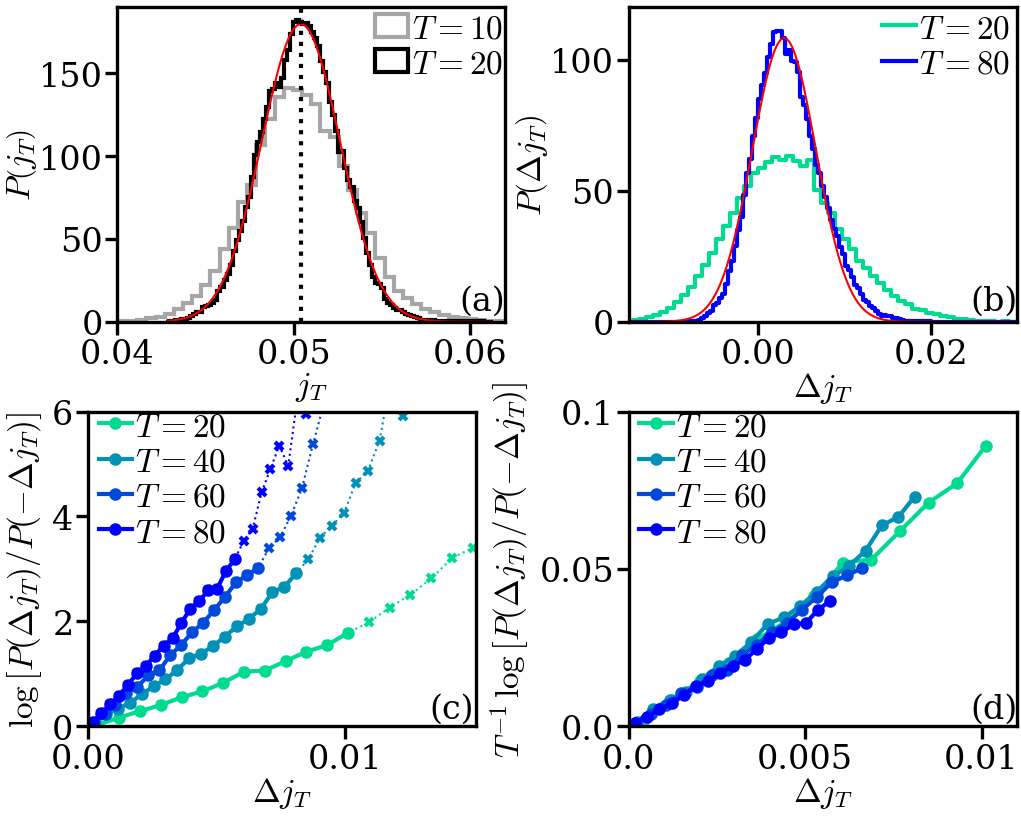}
\caption{\textbf{The GCEM fluctuation symmetry:} (a) A normalized histogram, suitably binned, of the time averaged energy flux $j_T$ over periods of length $T$, drawn by the motor in the reference state $\mathcal{R}$. The smooth solid red line is a Gaussian fit to the histogram for $T=20$. (b) The histogram $P(\Delta j_T)$ of the empirical energy flux into locomotive modes $\Delta j_T=j_T-\langle j_{T}\rangle_R$ for the \ball{} confined inside the disk arena $\mathcal{D}$ in Fig.~\ref{fig:ballIntro}b. The smooth solid magenta line is a Gaussian fit to the histogram for $T=80$, which shows deviations near the tails. (c) The variation of $\log \frac{P(\Delta j_T)}{P(-\Delta j_T)}$ with $\Delta j_T$ for various values of $T$ and (d) shows the scaling collapse confirming linear dependence on $T$.}
\label{fig:Distribution}
\end{figure}

This idea for quantifying BTRS using fluctuation relations has been theoretically exploited \cite{Fodor2016PRL,OByrne2022a} in the active matter context, but in terms of GCEM relation for entropy production involving high dimensional statistics of individual microscopic trajectories, which are hard to measure experimentally. In comparison, the parameter $s$ in the relation \eqref{eq:gc s} for the current statistics offers a simpler, experimentally accessible measure for BTRS. However, it is not a priori obvious whether the relation in \eqref{eq:gc s} is at all satisfied for the multi-modal internal energetics of an active particle sourced from chemical processes, where the notion of temperature is not evident. It is even less evident for the mechanical \ball{}, where the motor exerts a torque on the ball via the rotor to generate locomotion. 

To our surprise, we found a clear evidence for \eqref{eq:gc s} in the \ball{}. To observe this, it is essential to filter the energy current into the relevant locomotive modes that drive the persistent motion of the \ball{}, which is responsible for the BTRS seen from its trajectory. Only a portion of the energy from the motor goes into the locomotive degrees of freedom, while the rest is dissipated into other mechanical modes. The energy expenditures into dissipative modes are measured by decoupling the rotor from the shell of the ball, rendering the \ball{} immobile. We define this as the reference state $\mathcal{R}$.

As the motor rotates, it consumes energy, even in the reference state. The energy flux is measured by $j=I^2 R$, where $I$ is the electric current drawn by the motor and $R$ is its resistance. The empirical current $j_T$ in \eqref{eq:gc s} is the arithmetic mean of $j$ over a time window of length $T$. The $j_T$-values from different time windows fluctuate around an ensemble average value $\langle j_{T}\rangle$. A normalized histogram $P(j_T)$ is generated by sampling $j_T$ through sliding a fixed-length time window $T$ over a long time series of $j$. The distribution $P(j_T)$ in the reference state $\mathcal{R}$, shown in Fig.~\ref{fig:Distribution}a, becomes narrower for larger $T$. The peak of the distribution at $\langle j_T \rangle_\mathcal{R}$ represents the average current the motor draws in the reference state $\mathcal{R}$, which is lost into dissipative modes. Remarkably, despite non-zero dissipation, the distribution is Gaussian.

When the stator is rigidly coupled to the shell of the \ball{}, the torque exerted by the motor causes the ball to move. In this state, the $\Delta j_T=j_T-\langle j_T \rangle_R$ is the net energy flux that goes into the locomotive degrees of freedom of the active particle. The distribution $P(\Delta j_T)$ for two values of $T$ is shown in Fig.~\ref{fig:Distribution}b. Unlike in the reference state, the distribution is non-Gaussian and positively skewed with a non-zero mean $\langle \Delta j_T\rangle $. Such skewed distributions are  characteristics of non-equilibrium fluctuations \cite{Takeuchi2011a}.

Our astounding observation is that the distribution $P(\Delta j_T)$ follows the symmetry \eqref{eq:gc s}. To demonstrate this, we plot the logarithm of the ratio $\frac{P(\Delta j_T)}{P(-\Delta j_T)}$ for different sets of $T$ in Fig.~\ref{fig:Distribution}c. The individual plots confirm a linear growth with $\Delta j_T$ in accordance with the GCEM relation \eqref{eq:gc s}. The linear dependence extends to values of $\Delta j_T$ well beyond typical fluctuations, particularly for negative values, suggesting that the linearity is not merely a leading-order behavior of a more complex functional dependence. (Deviations from linearity appear at large fluctuations of $\Delta j_T$, where statistics are exponentially rare.) Furthermore, the linear dependence of the exponential on $T$ in \eqref{eq:gc s} is confirmed by the scaling collapse of $\frac{1}{T}\log\frac{P(\Delta j_T)}{P(-\Delta j_T)}$ for different $T$, as shown in Fig.~\ref{fig:Distribution}d. The collapsed data fits a straight line passing through the origin, presenting a convincing agreement with the GCEM-relation in \eqref{eq:gc s} (see \cite{Ciliberto2010JSM,Menon_2004} for comparison).

We stress that our experimental system is not an idealized instance of the GCEM fluctuation relation. There are no immediate Markovian descriptions of the dynamics, nor does $s j_T$ have an interpretation as the entropy production rate, unlike in thermal systems.

A point to note is that in \eqref{eq:gc s}, whether or not $j_T$ is made dimensionless by suitably normalizing with, say, $\langle j_T \rangle_\mathcal{R}$, does not change the value of $s$. Ideally, the parameter $s$ should be dimensionless, which for thermal systems is achieved by re-scaling with an effective energy scale \cite{Ciliberto2010JSM,Menon_2004}. Such an energy scale is ambiguous in our \ball{}. Nevertheless, the re-scaling does not change our following conclusions about the impact of kinematic constraint on the BTRS.

\subsection*{Effect of constraint}
\begin{figure}[t]
\centering
\includegraphics{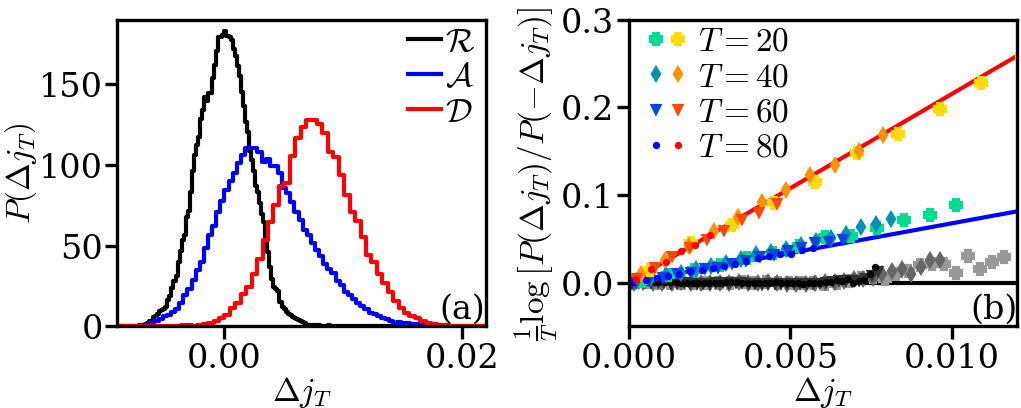}
\caption{\textbf{Effect of constraints:} (a) Comparison of $P(\Delta j_T)$ in the reference state $\mathcal{R}$ for $T=175$ and the two confining geometries $\mathcal{D}$ and $\mathcal{A}$ for $T=80$. (b) Scaling collapse of $\frac{1}{T}\log \frac{P(\Delta j_T)}{P(-\Delta j_T)}$ for different sets of $T$ for all three constraints $\mathcal{R}$, $\mathcal{D}$, and $\mathcal{A}$, following the color code in (a). For $\mathcal{R}$, where data for larger $T$ are available, the symbols cross, rhombus, triangle, circle are for $T = \{100,125,150,175\}$. For the solid black line the slope $s=0$, for the blue line $s \approx 6.77$, for the red line $s \approx 21.568$.}
\label{fig:Entropy}
\end{figure}

We perform our measurements of $s$ for two constrained geometries: (1) the \ball{} in a disc arena $\mathcal{D}$ and (2) in an annular arena $\mathcal{A}$, as illustrated in Fig.~\ref{fig:ballIntro}b and Fig.~\ref{fig:ballIntro}d. A comparison of their respective $P(\Delta j_T)$ is shown in Fig.~\ref{fig:Entropy}a. The distribution in the reference state $\mathcal{R}$ is shown for comparison. The effect of the constraints is clearly visible in the relative order of the peak positions, which indicates that $\mathcal{D}$ is the most irreversible scenario. This is quantitatively confirmed by the value of $s$ measured from the slope of the collapsed data shown in Fig.~\ref{fig:Entropy}b. The result $s(\mathcal{D})>s(\mathcal{A})>s(\mathcal{R})=0$ shows that stronger kinematic constraints in the annulus bring the system closer to equilibrium. Similar effects of constraints were seen in the GCEM relation for a granular medium \cite{Menon_2004}, where increasing the number of interacting particles reduces the temperature differences of effective baths.

\section{\ball{} in a passive medium}\label{sec:passive medium}
\begin{figure*}
\centering
\includegraphics[]{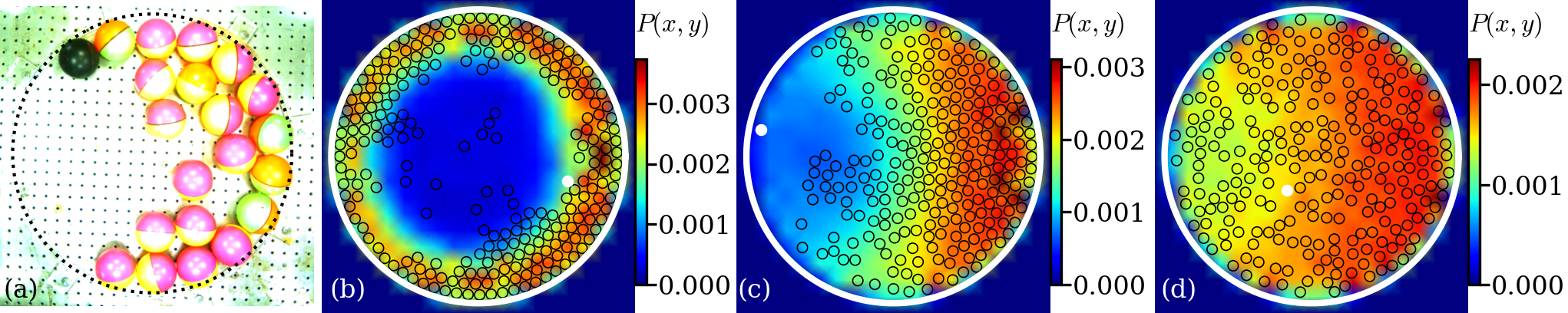}
\caption{\textbf{Active tracer in a passive medium:} (a) A configuration snapshot showing herding of passive \ball{}s (pink-yellow) by a single active \ball{} (black) confined inside the disc arena $\mathcal{D}$ indicated by the dotted circle.  (b-d) Typical configurations of one active Brownian particle (white) in a bath of passive particles following the under-damped dynamics \eqref{E:ABP}. The heat map with values indicated in the color bar shows the ensemble averaged density distribution of particles in a rotated coordinate system where the centre of mass is always placed at zero angle. (b) For high damping ($\gamma = 1.0$), the passive particles accumulate isotropically at the boundary of the disc arena. (c) For intermediate damping ($\gamma = 0.1$) the passive particles are herded in one half of the arena. (d) For low damping ($\gamma = 0.01$), passive particles are homogeneously distributed over the whole arena.}
\label{F:Herding}
\end{figure*}
BTRS brings the system outside the purview of equilibrium-constraints of entropy growth, allowing for new states of matter. A well-known example is  motility-induced phase separation \cite{Tailleur_Cates}, a phenomenon intensely studied in overdamped self-propelled particle models. These models, with purely repulsive steric interactions, undergo a macroscopic phase separation into a gas and some other phase of higher density \cite{Digregorio_2018,Klamser2018}.

To see what simple many-body self-organization the BTRS in the \ball{} can bring about, we confine a single active \ball{} together with many passive balls in the disk arena. The passive balls are unpowered \ball{}s that move due to momentum transfer during collisions with the \ball{}. This situation exemplifies an active tracer \cite{Kumar2011} in a passive medium.

We observe that the active \ball{} `herds' the passive balls in one side of the arena. A typical configuration is shown in Fig.~\ref{F:Herding}a. As shown in the movie in the Supporting Informations \cite{SI}, the `herding' is a non-equilibrium dynamical phenomenon: the crowd of passive balls does not stay at the same place in the arena, but dissolves and regroups at new positions.

To shed light on the mechanism behind this herding, we perform numerical simulations of one active Brownian particle in a bath of $(N-1)$ passive particles. We capture the geometry of the experiment by confining the particle positions $\mathbf{r}_i(t) = r_i(\cos\alpha_i,\sin\alpha_i)$ within a circular wall of radius $R$, modeled via a short-ranged repulsive potential $V(R-r_i)$ (see \cite{SI}). Inter-particle interaction is modeled via a short-range repulsive potential $U(r_{ij})=4 \varepsilon\left[(\sigma / r_{ij})^{12}-(\sigma / r_{ij})^{6}\right]+\varepsilon$ for the inter-particle distance $r_{ij} = |\mathbf{r}_i-\mathbf{r}_j|<2^{1 / 6} \sigma$ and $0$ otherwise. Here, $\sigma$ is a measure of the particle size and $\epsilon$ of the interaction strength. 

The net steric force acting on particle $i$ is $\mathbf{F}_i = - \sum_{j\ne i} \mathbf{\nabla}_i U(r_{ij}) - \mathbf{\nabla}_i V(R-r_i)$. The active Brownian particle is self-propelled by an active force $\mathbf{f}_t = f \mathbf{n}(\theta_t)$ of constant amplitude $f$ and with an orientation $\mathbf{n}(\theta_t) = (\cos\theta_t, \sin\theta_t)$ that diffuses in time as $\dot{\theta}_t = \eta_t$, where $\eta_t$ is a Gaussian white noise of zero mean and covariance $\left\langle\eta_t \eta_{t^{\prime}}\right\rangle=2 \tau_p^{-1} \delta\left(t-t^{\prime}\right)$, with $\tau_p$ being the persistence time. The position of the particles evolve by 
\begin{equation}
m\ddot{\mathbf{r}}_{i}(t) = -\gamma \dot{\mathbf{r}}_{i}(t) + \mathbf{F}_i + \begin{cases}
f \mathbf{n}(\theta_t),\, & \text{if } i \text{ is active,}\\
0, & \text{ otherwise,}
\end{cases}
\label{E:ABP}
\end{equation}
with the particle mass $m$ and the viscous damping $\gamma$. We set all parameters ($m$, $\varepsilon$, $\tau_p$, $f$) to unity, except for the damping $\gamma$, without loss of generality. Furthermore, we chose a packing fraction $\phi = (N\sigma^2)/(4R^2) = 0.08$, comparable to the experiment. We selected a rather large number of particles, $N = 256$, compared to the experiment to emphasise that the physics at play is scalable. The relevant parameter is therefore the damping $\gamma$, whose value determines whether we see the herding effect or not.

For large damping, the momentum transferred from the active particle to the passive particles is rapidly dissipated. This leads to a state where the active particle gently pushes most passive particles to the boundary, resulting in an angularly isotropic arrangement (see Fig.~\ref{F:Herding}b)).

At intermediate damping, the momentum can propagate via collision chains from the active particle to many passive particles. However, the rate at which kinetic energy is lost due to damping scales with the number of particles to which the energy is distributed (see \eqref{E:ABP}). Therefore, high-density regions of passive particles act as a momentum sink, while the active particle is the only source of momentum. Both the position of the sink and source are mobile and correlated with each other. As shown in Fig.~\ref{F:Herding}c, the passive particles accumulate, on average, in the half-arena opposite to the position of the active particle. While the active particle explores the arena, it `heats up' the area around itself through collision-induced momentum transfer. Due to the circular shape of the arena, the activated passive particles are bound to eventually travel towards the momentum sink, where collision-chains rapidly dissipate the energy. The position of the sink moves when the active particle tries to penetrate into the high-density region, giving the overall impression of herding (see movie in \cite{SI}).

When the damping is weak, the kinetic energy of a passive particle does not decay fast enough before it is reactivated by collision. This results in a state where all particles are constantly moving, homogenising the density distribution across the entire arena (see Fig.~\ref{F:Herding}d).

Based on the numerical observations, we conjecture that the herding effect observed in the experiment is a generic feature of a moving energy source within a bath of passive particles subject to moderate damping. For example, a Brownian particle at some finite temperature would also generate comparable density distributions in a passive medium upon varying the damping. The difference between this and a persistently moving energy source might be detectable through a careful study of the cluster dynamics of the herded state, which, however, lies outside the scope of this work.

From the mechanism of momentum dissipation via cascading collisions, it is evident that the damping $\gamma$ must be scaled with the system size for the herding effect. For example, a collision chain that is just the right length to generate the `herded' state for $N = 256$ particles will be insufficient to `herd' $N = 1024$ particles. Furthermore, the geometry of the arena impacts the phenomenon. While an elliptic arena leads to more efficient herding in one half, a square-shaped arena blurs this effect.

\section{Discussions}\label{sec:Conclusion}
The present work introduced a novel, controllable experimental system of active dynamics, consisting of a rolling sphere with an asymmetric mass distribution that operates by drawing energy from an internal battery. An unprecedented advantage of our system is the direct access to the energy distribution in internal modes, which allows us to explore thermodynamic principles in the energetics of individual active particles.

We demonstrated that the extended nature of the \ball{} requires consideration of its interaction with boundaries and with other \ball{}s, which plays a vital role in its non-equilibrium dynamics. To analyze BTRS, we introduced a novel, non-invasive measure based on collision events that are easy to capture in experiments. Relying on this irreversibility measure and by analyzing the system's energy dissipation under different kinematic constraints, we showed that BTRS is not only an inherent property of the active particle, but it also depends on environmental constraints. Further energetic insight came from our discovery of a non-trivial fluctuation symmetry in the internal modes of the active particle, which is expected to be universal. 

Several comments in order. Our observations indicate that an active system could be brought closer to equilibrium by imposing appropriate constraints. It would be interesting to investigate whether such constraints could be self-induced by interactions with many such active particles, potentially leading to an effective equilibrium regime for moderate activity, as noticed in theoretical studies \cite{Fodor2016PRL,OByrne2022a}.

Even though the \ball{} is an athermal system, the observation of the GCEM fluctuation relation provides an effective temperature $T_{\textrm{eff}}$ for the \ball{}, using $s=k_B^{-1}\left(\frac{1}{ T_\textrm{amb}}-\frac{1}{T_\textrm{eff}}\right)$, where $T_\textrm{amb}$ is the ambient temperature. The $T_\textrm{eff}$ bears no relation to the temperature of the battery, but instead changes with the environmental constraints. This concept of an effective temperature $T_{\textrm{eff}}$ for the many-particle case would be in line with the questions of thermodynamic state variables for active matter \cite{2015_Solon}.

Overall, the \ball{} system holds great promise as a prototype for studying non-equilibrium statistical physics, particularly in situations where a detailed understanding of energetics is required. The combination of our novel measures of BTRS and precise, controllable access to energy consumption of internal modes offers new tools for understanding active matter thermodynamics and discovering novel design principles for programmable active matter. This approach could help us understand how dissipation dictates structure and function in living matter.

\matmethods{

\subsection*{Mechanical Construction}
The experimental system, based on a Squiggle-ball toy, features an $80$mm diameter plastic spherical shell with a DC motor mounted along the axis. The motor's shaft connects to the shell, and the motor body is housed in an acrylic enclosure. A $10\times20\times40$mm, $57$g brass block is screwed to one side of the motor. A 2xAA battery holder and a PCB are attached to the adjacent sides of the acrylic enclosure. The asymmetric mass distribution causes the center of mass to shift as the motor rotates, making the sphere roll chaotically on a solid surface. When it collides with a rigid object, it rebounds randomly.

 \subsection*{Electronics}

 The electronics circuit schematic is shown in the Supporting Informations \cite{SI}. An ESP-12F micro-controller, based on the ESP8266 SoC, is used for control, data collection, and sensing. It features a 32-bit RISC processor running at 80 MHz, 36 KB RAM, 4 MB SPI flash, 10-bit ADC, I/O, and a PCB antenna. The MPU-6050 IMU measures angular velocity and linear acceleration, with gyroscope ranges of $\pm 250^{\circ}/\rm{s}$, $\pm 500^{\circ}/\rm{s}$, $\pm 1000^{\circ}/\rm{s}$, and $\pm 2000^{\circ}/\rm{s}$, and accelerometer ranges of $\pm 2\rm{g}$, $\pm 4\rm{g}$, $\pm 8\rm{g}$, and $\pm 16\rm{g}$. We use $\pm 2000^{\circ}/\rm{s}$ and $\pm 2\rm{g}$ ranges. Communication between MPU-6050 and ESP-12F is via $\rm{I}^2\rm{C}$, with data collected at 25 Hz and transmitted over Wi-Fi.
 The ESP8266's GPIO pins supply up to 6 mA, insufficient for the motor, so a DRV8833 dual H-bridge driver is used. A Hall effect current sensor, WCS2702, measures motor current, powered by a 3.3 V LT1963A regulator, with 2.3 mA resolution.
Any 3-6 V DC motor with a $10\times12~\rm{mm}$ cross section and $3~\rm{mm}$ D-shaped gearbox shaft can be used. For this experiment, a 60 rpm motor with a $9~\rm{mm}$ shaft was used, with no load current of $10~\rm{mA}$ and stall current of $1~\rm{A}$, and rated/stall torque of $200/1600~\rm{gm\text{-}cm}$.
 Two 3.7 V 800 mAh Li-ion batteries power the system: one for the motor and driver, the other for the micro-controller and electronics. Ground pins of all components are connected together.  Calibration procedures for the current sensor, gyroscope, and accelerometer are provided in the Supporting Informations \cite{SI}.

\subsection*{Measurements}

We take images of the ball from the top of the arena where it is moving and by detecting the position, we can calculate its linear velocity $\vec{v}$.  The angular velocity of the ball with respect to the body frame was computed from the gyroscopic data.

\subsection*{Software}
The source code for controlling the micro-controller, collecting data from the IMU and current sensors, and transmitting data over Wi-Fi is written in the Arduino IDE. After booting, the ESP-12F connects to a local Wi-Fi network using predefined SSID and password. It starts an HTTP web server on port 80 and a Telnet server on port 23. The web server is used to: (i) start measurements, (ii) toggle the GPIO2 pin to control the motor, (iii) upload code to the micro-controller (OTA updates), and (iv) reboot the micro-controller. When instructed to start measurements, the ESP-12F sets the GPIO2 pin to High, turning on the motor and collecting data from the IMU and current sensors. This data is sent to a computer on the same network via Telnet at $25~\rm{Hz}$.

Python scripts are used to collect sensor data over Wi-Fi and image data from a camera over USB. Synchronization of sensor and image data is achieved using the multirun plugin in PyCharm IDE, maintaining a sync within 20 seconds over 2 hours and 15 minutes. For simultaneous data collection from multiple "Squigglebots," ThreadPoolExecutor from the concurrent.futures module is used. All source codes are available on GitHub \cite{squigglebot_2022}.

}

\showmatmethods{} 

\acknow{SD, TS, and SG acknowledge the financial support of the Department of Atomic Energy, Government of India, under Project Identification No. RTI 4002.}

\showacknow{} 

\bibliography{Refs,roll_ball}

\end{document}